\documentclass[namedreferences]{solarphysics}
\usepackage[optionalrh,solaromanenum,natbib]{spr-sola-addons}
\usepackage{color}                       % For color text: \color command
\usepackage{url}                         % For breaking URLs easily trough lines
\usepackage{amsthm,amscd,amssymb,pxfonts}
\usepackage{latexsym}
\usepackage{graphicx,epsfig}
\usepackage[pdfborder={0 0 0},urlcolor=blue,colorlinks=true,citecolor=black,breaklinks=true]{hyperref}
\newcommand{\rfr}[1]{Equation (\ref{#1})}

\newcommand{\dert}[2]{\frac{{{\mathrm{d}}}{#1}}{{{\mathrm{d}}}{#2}}}
\newcommand{\virg}[1]{``#1''}
\newcommand{\bb}[3]{\bibitem[\protect\citeauthoryear{#1}{#2}]{#3}}
\newcommand{\eqi}{\begin{equation}}
\newcommand{\eqf}{\end{equation}}
\newcommand{\Om}{\mathrm{\Omega}}
\newcommand{\rp}[2]{{#1\over#2}}
\newcommand{\lb}[1]{\label{#1}}
\newcommand{\bds}[1]{\mathbfit{#1}}
\newcommand{\kap}{\hat{\bds k}}
\newcommand{\kx}{\hat{k}_x}
\newcommand{\ky}{\hat{k}_y}
\newcommand{\kz}{\hat{k}_z}

\newcommand{\doiurl}[1]{\href{http://dx.doi.org/#1}{\textsf{#1}}}
\newcommand{\adsurl}[1]{\href{http://adsabs.harvard.edu/abs/#1}{\textsf{#1}}}

\begin{document}

\begin{article}

\begin{opening}

\title{Constraining the Angular Momentum of the Sun with Planetary Orbital Motions and General Relativity}

\author{L.~\surname{Iorio}$^{1,2,3}$
}

\runningauthor{L. Iorio}
\runningtitle{General R
elativity and The Angular Momentum of the Sun}

\institute{$^{1}$Ministero dell'Istruzione, dell'Universit\`{a} e della Ricerca (M.I.U.R.)-Istruzione.
                     email: \href{mailto:lorenzo.iorio@libero.it}{\textsf{lorenzo.iorio@libero.it}}\\
             $^{2}$International Institute for Theoretical Physics and Advanced Mathematics Einstein-Galilei.
             email: \href{mailto:lorenzo.iorio@libero.it}{\textsf{lorenzo.iorio@libero.it}}\\
             $^{3}$ Fellow of the Royal Astronomical Society (F.R.A.S.). Viale Unit\`{a} di Italia 68, 70125, Bari, Italy.
             email: \href{mailto:lorenzo.iorio@libero.it}{\textsf{lorenzo.iorio@libero.it}}\\
             }

% Permanent address for correspondence: Viale Unit\`{a} di Italia 68, 70125, Bari (BA), Italy}

\begin{abstract}
The angular momentum of a star is an important astrophysical quantity related to its internal structure, formation, and evolution. Helioseismology yields $S_{\odot}= 1.92\times 10^{41}\ {\rm kg\ m^2\ s^{-1}}$ for the angular momentum of the Sun. We show how it should be possible to constrain it  in a near future by using the gravitomagnetic Lense--Thirring effect predicted by General Relativity for the orbit of a test particle moving around a central rotating body. We also discuss the present-day situation in view of the latest determinations of the supplementary perihelion precession $\Delta\dot\varpiup_{\mercury}$ of Mercury.
A fit by \citeauthor{fienga011} (\textit{Celestial Mech. Dynamical Astron.}{} \textbf{111}, 363, \citeyear{fienga011}) of the dynamical models of several standard forces acting on the planets of the solar system  to a long data record yielded $\Delta\dot\varpiup_{\mercury}^{\rm (meas)}=0.4\pm 0.6$ milliarcseconds per century. The modeled forces did not include the Lense--Thirring effect itself, which is expected to be as large as $\dot\varpiup^{(\rm LT)}_{\mercury} = -2.0\ {\rm mas\ cty}^{-1}$  from helioseismology-based values of $S_{\odot}$. By assuming the validity of General Relativity, from its theoretical prediction for the gravitomagnetic perihelion precession of Mercury one can straightforwardly infer
$S_{\odot}\leq 0.95\times 10^{41}\ {\rm kg\ m^2\ s^{-1}}$. It disagrees with the currently available values from helioseismology.
%contrary to an analysis of some years ago based upon  the Venus perihelion and the  EPM ephemerides which did not show any disagreement.
Possible sources for the present discrepancy are examined. Given the current level of accuracy in the Mercury ephemerides, the gravitomagnetic force of the Sun should be included in their force models. MESSENGER, in orbit around Mercury since March 2011,  will collect science data until 2013, while \textit{BepiColombo}, to be launched in 2015, should reach Mercury in 2022 for a year-long science phase: the analysis of their data will be important in effectively constraining $S_{\odot}$ in about a decade or, perhaps, even less.
\end{abstract}

\keywords{Ephemeris$\cdot$Helioseismology$\cdot$Rotation}

\end{opening}
%-------------------------------------------------

\section{Introduction}
The angular momentum $\mathrm{[}S\mathrm{]}$ of a main-sequence star
can yield relevant information about its inner
properties and its activity \citep{antia00,yang06,yang08,bi011}. More specifically, $S$ is related to the stellar internal rotation rate \citep{pijpers98}. Moreover,
it is an important diagnostic for testing theories
of stellar formation \citep{pijpers03}. The angular momentum
can also play a decisive role in stellar evolution,
in particular towards  higher mass \citep{tarafdar71,wolff82,vigneron90,wolff97,dimauro00,komm03,herbst05,jackson05,antia08}.

%In passing, it may be interesting to recall that, in the case of a
%rotating \citet{kerr63} black hole of mass $M$, there is a
%theoretical upper limit
%\eqi S^{\rm (max)} =\rp{M^2 G}{c}, \eqf
%where $G$ is the Newtonian constant of gravitation
%and $c$ is the speed of light in vacuum, so that \citep{shapiro86}
%\eqi S = \chi_g\ S^{\rm (max)},\ \left|\chi_g\right| \leq 1.\eqf
%If $\left|\chi_g\right| > 1$, a naked singularity without a horizon
%would occur, along with the possibility of causality
%violations because of closed timelike curves \citep{chandrasekhar83}. Incidentally, we remark that, although not
%yet proven, the cosmic censorship conjecture \citep{penrose69}
%states that naked singularities cannot be formed
%via the gravitational collapse of a body.
%The case of compact stars was recently treated
%by \citet{lo011}, showing that for neutron stars with $M\gtrsim 1 M_{\odot}$
%it should be \eqi \chi_g \lesssim 0.7,\eqf independently of the Equation Of
%State (EOS) governing the stellar matter. Hypothetical
%quark stars may have \eqi \chi_g > 1,\eqf strongly depending on the
%EOS and the stellar mass \citep{lo011}.
%
%On the
%contrary, it is known that  $\chi_g$ of main-sequence stars depends in
%a non-negligible way on the stellar mass.  Thus, the dimensionless spin
%parameter can well be  \citep{kraft69,kraft70,dicke70,gray82} \eqi \left|\chi_g\right|\gg 1;\eqf for example,
%from the analysis by \citet{iorio010} it can be inferred
%that \eqi \chi_g \approx 36\eqf for the star HD15082 (WASP-33) \citep{grenier99} having $M=1.495 M_{\odot}$ \citep{collier010}.

In this work, we propose to put  dynamical constraints on $S_{\odot}$, independent of any model of the Sun's interior,  by exploiting General Relativity. See also \citet{Iorio08} for earlier investigations. The plan of the article is as follows: In Section \ref{conoscenza} we summarize the current level of knowledge of the solar angular momentum from other, non-dynamical techniques. In Section \ref{grg} we introduce the general-relativistic Lense--Thirring effect, which is related to the angular momentum. We apply it in Section \ref{pianeti} in connection with the latest determinations of the orbital motions
of the inner planets of the solar system \citep{fienga011}. In Section \ref{spiego} we discuss the impact that other competing dynamical effects may have. Section \ref{discussione} summarizes our findings.
\section{The Angular Momentum of the Sun: Present-Day Knowledge}\lb{conoscenza}
The angular momentum is one of some global properties of the Sun of astrophysical interest, which  are related to the internal rotation rate through integral
equations \citep{pijpers98}. The solar internal
rotation rate can be determined using inverse techniques of helioseismology \citep{dalsgaard03,kosovichev03,dimauro03,dimauro08} applied to data collected from both the ground-based \textit{Global Oscillation Network Group} \citep[GONG:][]{harvey96}
and  the instruments carried onboard the \textit{SOlar and Heliospheric Observatory} \citep[SOHO:][]{domingo95}, such as the
\textit{Solar Oscillations Investigation/Michelson Doppler Imager} \citep[SOI/MDI:][]{scherrer95} and \textit{Global Oscillations at Low Frequencies}  \citep[GOLF:][]{gabriel95}, and the \textit{Solar Dynamics Observatory/Helioseismic and Magnetic Imager} \citep[SDO/HMI:][]{Scherr012}.

Such observations pertain the  solar oscillation frequencies.  The observed solar oscillations correspond to standing
acoustic waves maintained by pressure forces, which form
the class of the $p$-modes, and to standing surface gravity
waves, maintained by gravity, known as $f$-modes. In addition,
one should mention the probable, although quite discussed,
existence of the internal gravity waves, $g$-modes, which are
sensitive to the structure and rotation of the deeper interior of
the Sun \citep{dimauro08}.
Oscillations have several advantages over all the other observables:
their frequencies  can be measured with
high accuracy, and depend in a quite simple way on the equilibrium
structure of the Sun \citep{dimauro08}.

Table \ref{tavola} summarizes the present-day knowledge of $S_{\odot}$ according to mainly helioseismology.
\begin{table}
\caption{Values of the solar angular momentum, in units of $10^{41}\ {\rm kg\ m^2\ s^{-1}}$, retrieved from the literature. The uncertainties, when released by the authors, have been quoted as well. With the exception of the figure by \citet{livingstone00}, inferred from surface rotation, all the other values quoted come from helioseismology. \citet{pijpers98} used data from both GONG and SOHO/MDI. \citet{dimauro00} relied upon MDI and GOLF. \citet{komm03} used GONG and MDI data as well, but in a different way with respect to \citet{pijpers98} and \citet{antia00}. The values by \citet{yang06,yang08} and \citet{bi011} were obtained from models taking into account the effects of the magnetic field: we quoted only those figures consistent at $1-3\sigma$ level with helioseismological results.
}\label{tavola}
\begin{tabular}{ccc}
\hline
$S_{\odot}\ [10^{41}\ {\rm kg\ m^2\ s^{-1}}]$ & $\sigma_{S_{\odot}}\ [10^{41}\ {\rm kg\ m^2\ s^{-1}}]$ & Reference \\
\hline
$1.90$ & $0.015$ & \citep{pijpers98} \\
$1.63$ & $-$ & \citep{livingstone00}\\
$2.02$ & $0.04$ & \citep{dimauro00}\\
$1.91$ & $-$ & \citep{antia00} \\
$1.94$ & $0.05$ & \citep{komm03} \\
$1.91$ & $-$ & \citep{yang06}  \\
$2.045$ & $-$ & \citep{yang08}  \\
$2.02$ & $-$ & \citep{bi011}  \\
\hline
\end{tabular}
\end{table}
On average, it is \eqi S_{\odot} = 1.92\times 10^{41}\ {\rm kg\ m^2\ s^{-1}}.\eqf
%Since for the Sun
%\eqi \rp{M^2_{\odot}G}{c}=8.80\times 10^{41}\ {\rm kg\ m^2\ s^{-1}},\eqf it follows \eqi \chi_g^{\odot} = 0.21.\eqf
%
\section{Using General Relativity: the Lense--Thirring Effect}\lb{grg}
Contrary to the Newtonian theory of gravitation, the Einsteinian General Relativity predicts the existence of  dynamical effects which can, in principle, be used to
measure, or, at least, constrain the angular momentum of a  rotating body since their analytical expressions are directly proportional to $S$.

The exterior spacetime metric of a typical astronomical object can adequately
be described within the Parameterized Post-Newtonian (PPN) approximation \citep{soffel89}. Within such a framework, a stationary gravitomagnetic field
 arises around a slowly rotating mass with proper angular
momentum $S$ \citep{thorne86,thorne88,mashoon01}.  Gravitomagnetism does not refer to any combined effect of electromagnetism and gravitation.
%At great distance $r$ from the spinning body, General Relativity predicts that its gravitomagnetic field is  \citep{thorne86,thorne88,mashoon01}
%
%
%\eqi\bds B_g=-\rp{G}{cr^3}\left[\bds S-3\left(\bds S\bds\cdot\bds{\hat{r}}\right)\bds{\hat{r}}\right].\lb{bigi}\eqf
%
%
%$\bds B_g$
Such a purely formal denomination is due to the fact that the gravitomagnetic field affects a
test particle moving with velocity $\bds v$ with a noncentral,
Lorentz-like acceleration
%\footnote{The general-relativistic multiplicative factor 2 in \rfr{alt} corresponds to $1+\gamma$ in PPN formalism.}
\citep{mccarthy04}
%
%
%
%\eqi\bds A_{\rm LT}=-2\left(\rp{\bds v}{c}\right)\bds\times \bds B_g,\lb{alt}\eqf
%
%
which is analogous to the one felt by a moving
electric charge in a magnetic field in the framework
of  Maxwellian electromagnetism. It is the cause of the so-called Lense--Thirring effect \citep{lense18} consisting of small secular precession of  some of the osculating Keplerian orbital elements  of the orbit of a test particle.

The gravitomagnetic precession of a gyroscope \citep{pugh59,schiff60}
%, which is another theoretical consequence of \rfr{bigi},
was recently measured in a dedicated space-based experiment, known as \textit{Gravity Probe B} \citep[GP-B:][]{everitt74}, performed with four cryogenic gyroscopes carried onboard a drag-free spacecraft orbiting the Earth. The prediction of General Relativity for such an effect was successfully corroborated at a claimed $19\%$ accuracy \citep{everitt011}.
It may be interesting to recall that \citet{haas75} proposed to measure the same effect with a drag-free spacecraft orbiting the Sun.

In this article, we will show how to use the gravitomagnetic orbital precessions to constrain the solar angular momentum. Some spacecraft-based missions
were proposed in the more or recent past
to accurately measure the Sun's gravitomagnetic
field by means of its direct effects on the propagation
of electromagnetic waves. They are
the \textit{Laser Astrometric Test of Relativity} \citep[LATOR:][]{turyshev04}, which aimed to directly measure the
frame-dragging effect on the light with a $\approx 0.1\%$
accuracy \citep{turyshev09}, and the \textit{Astrodynamical Space
Test of Relativity using Optical Devices I}
\citep[ASTROD I:][]{ni08}, whose goal is to measure the gravitomagnetic component of the time delay with a
$10\%$ accuracy \citep{appourchaux09}. We notice that any tests of general-relativistic solar frame-dragging would necessarily be accurate to a percent level. Indeed, the angular momentum of the Sun should be considered as known independently of the Lense--Thirring effect itself; helioseismology yields results accurate just to that level, as shown by Table \ref{tavola}.

\citet{Iorio012} analytically worked out the secular Lense--Thirring precessions of the Keplerian orbital elements for a generic orientation of the angular momentum of the central body in a form that allows for easy comparisons with the observation-based quantities usually determined by the astronomers.
They are \citep{Iorio012}
\begin{equation}
\begin{array}{lll}
\left\langle\dert a t \right\rangle & = & 0, \\ \\
\left\langle\dert e t \right\rangle & = & 0, \\ \\
\left\langle\dert I t \right\rangle & = & \rp{ 2 G S \left( \kx \cos\Om + \ky \sin\Om \right) }{ c^2 a^3 (1-e^2)^{3/2} }, \\ \\
\left\langle\dert\Om t \right\rangle & = & \rp{ 2 G S \left[ \kz + \cot I\left(\ky \cos\Om - \kx \sin\Om\right)\right] }{ c^2 a^3 (1-e^2)^{3/2} }, \\ \\
\left\langle\dert\varpiup t \right\rangle & = & -\rp{ 2 G S \left\{2\left[\kz \cos I + \sin I \left( \kx \sin\Om - \ky \cos\Om \right)  \right]
-\left[ \kz \sin I + \cos I \left( \ky \cos\Om - \kx \sin\Om \right) \right]\tan(I/2)
\right\} }{ c^2 a^3 (1-e^2)^{3/2} }, \\ \\
\left\langle\dert{\mathcal{M}}t\right\rangle & = & 0;
\end{array}\lb{piccololt}
\end{equation}
where $\kap=\left(\kx,\ky,\kz\right)$ is the unit vector of the rotation axis of the central body in the reference frame adopted, while $a,e,I,\Om,\mathrm{\varpiup},$ and $\mathcal{M}$ are the semimajor axis, the eccentricity, the inclination, the longitude of the ascending node, the longitude of pericenter, and the mean anomaly, respectively, of the orbit of the test particle \citep{murray99}. In particular, $I$ is the inclination of the orbital plane to the reference $\left(x,y\right)$ plane, $\Om$ is an angle in the reference $\left(x,y\right)$ plane counted from the reference $x$-direction to the line of the nodes which is the intersection of the orbital plane to the reference $\left(x,y\right)$ plane, and $\varpiup\doteq\Omega+\omegaup$ is a \virg{dogleg} angle since the argument of pericenter $\omegaup$ is an angle in the orbital plane counted from the line of the nodes to the point of closest approach \citep{murray99}. The precessions of \rfr{piccololt} are averages over one full orbital revolution of the test particle; they were obtained \citep{Iorio012} by using the Lagrange perturbative equations \citep{murray99} applied to a suitable perturbing potential \citep{Bar75} averaged over one orbital revolution. It should be noticed that almost all of the usual derivations of the Lense--Thirring effect  existing in
literature are based on the particular choice of aligning $\bds S$ along the $z$-axis. For some exceptions, in different contexts, see \citet{barker70} and \citet{will08}. However, the authors of such works either did not explicitly work out the precessions or adopted a different parameterization in such a way that their results could not straightforwardly be adapted to the present context.

%To this aim,
%it can be noted that \rfr{piccololt} yields just the usual Lense--Thirring rates for $\hat{k}_x=\hat{k}_y\rightarrow 0,\hat{k}_z\rightarrow 1$. Indeed,
%by posing $I/2\doteq \alpha$, it is easy to show that $-2\cos I + \sin I\tan (I/2)=1-3\cos I$.

In the case of the Sun, \citet{fienga011} adopted the mean Earth's Equator at the epoch J$2000.0$ for their analysis; in such a frame, the unit vector of the Sun's spin axis $\mathrm{[}\kap_{\odot}\mathrm{]}$, entering \rfr{piccololt}, is not aligned with the $z$-axis. Indeed, it turns out
\eqi
\begin{array}{lll}
\hat{k}^{\odot}_x &=& \cos\delta_{\odot}\cos\alpha_{\odot}= 0.122, \\ \\
\hat{k}^{\odot}_y &=& \cos\delta_{\odot}\sin\alpha_{\odot} =-0.423, \\ \\
\hat{k}^{\odot}_z &=&\sin\delta_{\odot}= 0.897,\lb{spinaxis}
\end{array}
\eqf
where \citep{seidelmann07}
\eqi
\begin{array}{lll}
\delta_{\odot} & = &   63.87\ {\rm deg}, \\ \\
 \alpha_{\odot} & = & 286.13\ {\rm deg},
\end{array}
\eqf
are the declination and the right ascension, respectively,
of the Sun's North Pole
of rotation with respect to the mean terrestrial
equator at J2000.0.
\section{Constraints from Planetary Orbital Motions}\lb{pianeti}
Recently,  \citet{fienga011} processed a huge amount of planetary observations of various kinds, covering almost one century (1914--2010), with the dynamical force models of the INPOP10a ephemerides \citep{preprint}. The data included, among other things, also high-quality Doppler range-rate observations to the \textit{MErcury Surface, Space ENvironment, GEochemistry, and Ranging}
spacecraft \citep[MESSENGER:][]{solomon07} collected\footnote{More precisely,  \citet{fienga011} neither used direct observations to MESSENGER to fit the INPOP10a ephemerides nor estimated the spacraft's orbit.
Instead, they used normal points extracted from the SPICE/NAIF MESSENGER orbit [\url{http://naif.jpl.nasa.gov/naif/}].
\citet{fienga011} plan to themselves analyze such data in the coming months.} during its three flybys of Mercury in  2008\,--\,2009 \citep{smith010}. Thus, our knowledge of the orbit of the innermost planet of the solar system has been greatly improved. Generally speaking, Mercury is a very difficult planet to observe from the Earth because of a number of reasons \citep{balogh07}. Also orbiting Mercury by spacecraft is a challenge, since the planet is deep inside the gravitational potential well of the Sun \citep{balogh07}. Moreover, a very hostile thermal environment awaits any spacecraft in Hermean
%
%\footnote{\url{http://en.wikipedia.org/wiki/List\_of\_adjectivals\_and\_demonyms\_of\_astronomical\_bodies\#cite\_note-lewis-4}}
%
orbits \citep{balogh07}. As a result, to date only one spacecraft, \textit{Mariner 10}, reached Mercury more than 30 years
ago. See \citet{balogh07} for a general overview on the missions to Mercury. MESSENGER was inserted into  orbit around Mercury on 17 March 2011 \citep{harris011} for a one-year-long, near-polar-orbital observational campaign \citep{solomon07}. A year-long extension until March 2013 was recently approved [\url{http://www.space.com/13655-mercury-spacecraft-messenger-mission-extension.html}]. \textit{BepiColombo} \citep{grard01,benkhoff010} is another planned mission to Mercury whose launch is currently planned for 2015 [\url{http://sci.esa.int/science-e/www/area/index.cfm?fareaid=30}], while its one-year mission around Mercury should start after its arrival scheduled for 2022. Within the framework of the
\textit{Mercury Orbiter Radioscience Experiment}
\citep[MORE:][]{iess09}, accurate measurement of its range should be able to notably improve the determination of Mercury's orbit as well \citep{iess07}.

Apart from usual Newtonian mechanics, including the Sun's quadrupole mass moment $J_2$ \citep{rozelot011} as well, the mathematical models of INPOP10a for the solar system dynamics  included also the first post-Newtonian (1PN) static, Schwarzschild-like component of the gravitational field of the Sun \citep{mccarthy04} expressed in terms of the usual Parameterized-Post-Newtonian (PPN) parameters $\gamma$ and $\beta$, which are equal to one in General Relativity. Instead, \citet{fienga011} did not model the gravitomagnetic field of the Sun.

\citet{fienga011} estimated, among other things,  corrections $\Delta\dot\Om^{\rm (meas)}$ and $\Delta\dot\varpiup^{\rm (meas)}$ to the standard Newtonian--Einsteinian  secular precessions of the longitudes  of  node $\Om$ and  perihelion $\varpiup$ for the first six planets. In principle, such corrections account for any unmodeled/mismodeled dynamical features of motion, so that they could be used to pose constraints on the magnitudes of putative anomalous effects, if any, with respect to the modeled ones. Concerning Mercury, they are \citep{fienga011}
\eqi
\begin{array}{lll}
\Delta\dot\Om^{\rm (meas)}_{\mercury} & = & 1.4\pm 1.8\ {\rm mas \ cty^{-1}}, \\ \\
\Delta\dot\varpiup^{\rm (meas)}_{\mercury} & = & 0.4\pm 0.6\ {\rm mas \ cty^{-1}}, \lb{correz}
\end{array}
\eqf
where mas cty$^{-1}$ is a shorthand for milliarcseconds per century.
The values of \rfr{correz} are statistically compatible with zero, so that they could be used, in principle, to put upper bounds on $S_{\odot}$ by comparing \rfr{piccololt} to \rfr{correz}. It makes sense since the predicted Lense--Thirring precession for it is larger than $0.6$ mas cty$^{-1}$, as it turns out
by using the values for $S_{\odot}$ listed in Table \ref{tavola}. In Table \ref{tavola2} we display the corresponding Lense--Thirring precessions for the perihelion of Mercury. On average, they yield \citep{Iorio08} \eqi \dot\varpiup^{\rm (LT)}_{\mercury} = -2.0\ {\rm mas\ cty}^{-1};\eqf departures from such a value are not relevant since they are smaller than $0.6$ mas cty$^{-1}$. \citet{Iorio08} used the usual formula with $\bds S$ parallel to the $z$-axis, but it turns out that, contrary to the node and the inclination, it does not affect the perihelion precession to a significant level of accuracy.
\begin{table}
\caption{Expected Lense--Thirring perihelion precession $\dot\varpiup_{\mercury}^{\rm (LT)}$ for Mercury, in mas cty$^{-1}$, according to \rfr{piccololt} computed with the figures of Table \ref{tavola} for the solar angular momentum. See also  \citep{Iorio08}. They are larger than the uncertainty released by \citet{fienga011} for the correction to the standard perihelion precession of Mercury, which amounts to $0.6$ mas cty$^{-1}$. We also show the discrepancy among the predicted $\dot\varpiup^{\rm (LT)}_{\mercury}$ and the determined $\Delta\dot\varpiup_{\mercury}^{\rm (meas)}$ of \rfr{correz} in number of $\sigma$.
}\label{tavola2}
\begin{tabular}{ccc}
\hline
$\dot\varpiup^{\rm (LT)}_{\mercury}\ {\rm (mas\ cty^{-1})}$ & Discrepancy (number of $\sigma$) & Reference for $S_{\odot}$\\
\hline
$-2.0$ & $4$ &  \citep{pijpers98} \\
$-1.7$ & $3.5$ & \citep{livingstone00}\\
$-2.1$ & $4.2$  & \citep{dimauro00}\\
$-2.0$ & $4$ & \citep{antia00} \\
$-2.0$ & $4$ & \citep{komm03} \\
$-2.0$ & $4$ & \citep{yang06}  \\
$-2.1$ & $4.2$ & \citep{yang08}  \\
$-2.1$ & $4.2$ & \citep{bi011}  \\
\hline
\end{tabular}
\end{table}
It can be noticed that there is a discrepancy of $4\sigma$, on average, among the predictions of General Relativity, computed using the helioseismically-inferred values for $S_{\odot}$, and the admissible range $-0.2\ {\rm mas \ cty^{-1}}\leq \Delta\dot\varpiup^{\rm (meas)}_{\mercury} \leq 1\ {\rm mas\ cty^{-1}}$ for unmodelled dynamical effects affecting the perihelion of Mercury obtained from observations by \citet{fienga011}.

We did not consider the Lense--Thirring node precessions $\dot\Om^{\rm (LT)}_{\mercury}$ since they are expected to be smaller than 1 mas cty$^{-1}$.

A direct comparison of \rfr{piccololt} to \rfr{correz} yields, in units of $10^{41}\ {\rm kg\ m^2\ s^{-1}}$,
\eqi S_{\odot}\leq 0.95.\eqf It is notably smaller that the figures of Table \ref{tavola}, at more than $10\sigma$ level.

In Section  \ref{spiego} we offer a preliminary discussion of possible explanations.

\section{Possible Explanations}\lb{spiego}
It must be stressed  that the previous considerations would be valid if the supplementary advance of the perihelion of Mercury of \rfr{correz} were \textit{entirely} explained in terms of frame-dragging.

Actually, in principle, there is the possibility that the unmodeled gravitomagnetic effect was partially or totally removed from the post-fit signature in the data-reduction process, having been somewhat \virg{absorbed} in the values of some of the standard parameters estimated in the fits like, \textit{e.g.}, the planetary initial conditions. After all, the expected magnitude of the Lense--Thirring effect is about of the same order of magnitude of the present-day accuracy in determining the orbit of Mercury. The PPN parameters $\beta$ and $\gamma$ entering the modeled static 1PN solar field were kept fixed to their general-relativistic values in the solutions yielding $\Delta\dot\varpiup^{\rm(meas)}$ and $\Delta\dot\Om^{\rm(meas)}$ \citep{fienga011}.

A possible conventional explanation in terms of orbital dynamics might  be the following. The corrections $\Delta\dot\varpiup$ were estimated by  \citet{fienga011} by modeling, among other things, the Newtonian action of the quadrupole mass moment $\mathrm{[}J_2\mathrm{]}$ of the Sun \citep{rozelot011} as well: more specifically,  $J_2^{\odot}$ was kept fixed to a reference value, which seems to be \citep{FiengaIAU} \eqi J_2^{\odot}=1.8\times 10^{-7}.\eqf For a historical overview of the impact of $J_2$ on  the measurement of the larger 1PN Schwarzschild-like perihelion precession of Mercury \citep[\textit{e.g.},][]{Pire03}.
Since such a physical quantity is  known with a necessarily limited accuracy, of the order of \citep{FiengaIAU,fienga011} $10\%$, the corrections $\Delta\dot\varpiup$ account, in principle, not only for the completely unmodeled Lense--Thirring effect, but also for the mismodeled precessions $\Delta\dot\varpiup^{J_2^{\odot}}$ due to the solar oblateness itself. Thus, a mutual cancelation might have occurred leaving just \rfr{correz}.
%
%
%
%
%
%Figure \ref{figura1} shows
To check this possibility, one has to consider the sum of the  precessions of the perihelion of Mercury caused by the Lense--Thirring effect and by a correction $\Delta J_2^{\odot}$ to the reference value used; analytical expressions for the long-term orbital precessions induced by the oblateness of the primary for an arbitrary direction of $\kap $ were recently obtained in a form suitable for comparison with observations in terms of usual osculating Keplerian orbital elements \citep{Iorio011b}.
%
%
%they are
%\eqi
%\begin{array}{lll}
%\dot\varpiup^{J_2} &=& \rp{3}{16}n\left(\rp{R}{a}\right)^2\rp{J_2}{\left(1-e^2\right)^2}\left\{ 8-11\kx^2-11\ky^2-2\kz^2 +\right.\\ \\
%
%&+&\left.\left(\kx^2+\ky^2-2\kz^2\right)\left(4\ci-5\cos 2 I\right)-\right.\\ \\
%
%&-& \left.4\left(\kx^2-\ky^2\right)\left(3+5\ci\right)\sin^2\left(\rp{I}{2}\right)\cos 2\Om-\right.\\ \\
%
%&-& \left. 2\ky\kz\sec\left(\rp{I}{2}\right)\left[\sin\left(\rp{3I}{2}\right) +5\sin\left(\rp{5I}{2}\right)\right]\cO +\right.\\ \\
%
%&+& \left. 2\kx\kz\sec\left(\rp{I}{2}\right)\left[\sin\left(\rp{3I}{2}\right) +5\sin\left(\rp{5I}{2}\right)\right]\sO -\right.\\ \\
%
%&-& \left. 8\kx\ky\sin^2\left(\rp{I}{2}\right)\left(3+5\ci\right)\sin 2\Om\right\},
%\lb{peri00}
%\end{array}
%\eqf
%
%
%where $n\doteq\sqrt{GM/a^3}$ is the Keplerian mean motion of the test particle, and $R$ is the equatorial mean radius of the central body.
%
%

Similar considerations hold also for the 1PN Schwarzschild-like component of the Sun's gravitational field: in the PPN formalism, the resulting perihelion precession for Mercury consists of the familiar $43.98$ arcsec cty$^{-1}$ rescaled by the multiplicative coefficient $(2 + 2\gamma - \beta)/3$ written in terms of the usual PPN parameters $\beta$ and $\gamma$. Actually, \citet{fienga011} modeled it  in the solution yielding \rfr{correz} by keeping $\beta$ and $\gamma$ fixed to one, but the current uncertainty in $\beta$, evaluated by \citet{fienga011} themselves as large as
\eqi\beta -1 = (-0.41\pm 0.78)\times 10^{-4}\lb{beta}\eqf
in a different solution by using the bounds on $\gamma$ from \citet{Ber03}, may translate into a further mismodeled competing perihelion precession $\Delta\dot\varpiup_{\mercury}^{\beta}$ which, added to $\Delta\dot\varpiup_{\mercury}^{J_2^{\odot}}$ and the expected $\dot\varpiup_{\mercury}^{\rm LT}$, may yield just \rfr{correz} from a mutual cancelation. It turns out that $\Delta\dot\varpiup_{\mercury}^{\beta}$ alone, computed with \rfr{beta} and added to $\dot\varpiup_{\mercury}^{\rm LT}$, is not able to reproduce \rfr{correz}. Instead, by including also $\Delta\dot\varpiup_{\mercury}^{J_2^{\odot}}$, computed from Equation (20)\,--\,Equation (21) of \citet{Iorio011b} with
\begin{equation} 0.1\times 10^{-7}\lesssim \Delta J_2^{\odot} \lesssim 0.2\times 10^{-7},\end{equation}
a range compatible with \rfr{correz} is obtained.

In principle, also the impact  of the oblateness of Mercury itself on its orbital motion should be considered, since it was not modeled in INPOP10a. However, it turns out that it is insufficient to cancel  the Lense--Thirring effect to the level of \rfr{correz}. Indeed, by using  Equation (20)--Equation (21) of \citet{Iorio011b}  with \citep{seidelmann07}
\eqi
\begin{array}{lll}
 R_{\mercury} & = & 2.4397\times 10^6\ {\rm m},\\ \\
 \alpha_{\mercury} & = & 281.01\ {\rm deg}, \\ \\
 \delta_{\mercury} & = & 61.45\ {\rm deg}, \\ \\
 \end{array}
 \eqf
 and \citep{smith010}
 \eqi J_2^{\mercury}  =  (1.92\pm 0.67)\times 10^{-5},\eqf
 the resulting perihelion precession is just
 \eqi\dot\varpiup^{J_2^{\mercury}}=(0.03\pm 0.01)\ {\rm mas\ cty^{-1}}.\eqf

In principle, viable candidates are also the major asteroids and the ring of the minor asteroids, although their dynamical action was accurately modeled by \citet{fienga011}. Indeed, both individual distant bodies and an external massive ring induce secular perihelion precessions.
Let us, first, consider the impact of the uncertainty in the mass of some of the individual major asteroids such as (1) Ceres, (2) Pallas and (4) Vesta.
%
%The secular perihelion precession of a planet with small orbital inclination due to a remote, pointlike object of mass $m_{\rm X}$ and distance $d_{\rm X}$ is %proportional to \citep{Iorio012b}
%\eqi\dot\varpiup^{\rm X}\approx \rp{24 Gm_{\rm X}}{128 n d_{\rm X}^3\sqrt{1-e^2}}.\lb{astero}\eqf
%
They are at about $2.4-2.8\ {\rm au}$.  The present-day uncertainty in their masses is \citep{luzum011}
\eqi \sigma_{m_{\rm ast}}=3\times 10^{-12}M_{\odot};\eqf \citet{fienga011} report for them similar or smaller uncertainties. Thus,
Equation (13)\,--\,Equation (14) of \citet{Iorio012b} yield
\eqi \Delta\dot\varpiup_{\mercury}^{\rm ast} \approx 1\times 10^{-3}\ {\rm mas\ cty^{-1}},\eqf which is, actually, of no concern. The secular perihelion precession induced by an external massive ring with mass $m_{\rm r}$ and radius $R_{\rm r}$ has been treated by a number of authors with different approaches \citep[\textit{e.g.},][]{Fienga08,kuchynka010,IorioEMB}.
%
%\eqi\dot\varpiup^{\rm r}  \approx \rp{3 Gm_{\rm r}\sqrt{1-e^2}}{4 n R_{\rm r}^3}.\lb{ringo}\eqf
%
For \citep{kuchynka09} \eqi R_{\rm r}=2.80\ {\rm AU},\ \sigma_{m_{\rm r}} = 2\times 10^{-11}M_{\odot},\eqf
they yield
\eqi \Delta\dot\varpiup_{\mercury}^{\rm r} \approx 0.02\ {\rm mas\ cty^{-1}}, \eqf
which is negligible as well since it is 30 times smaller that the accuracy in determining the supplementary perihelion advance of Mercury.

Finally, it is not unreasonable to argue that, ultimately, \rfr{correz} is based just on three normal points from MESSENGER, so that the entire matter should be left on hold until more data from the current Mercury orbiter will be gathered and analyzed.
\section{Summary and Conclusions}\lb{discussione}
We showed how it is possible, in principle, put constraints on the angular momentum of the Sun, independently of models of its interior, by exploiting the general-relativistic gravitomagnetic Lense--Thirring effect. It consists of secular precessions of the node and the pericenter of a test particle orbiting a central rotating body which are proportional just to its angular momentum.

Contrary to the 1PN gravitoelectric Schwarzschild-like component of the gravitational field of the Sun and all standard Newtonian effects, the solar gravitomagnetic field was not included in the mathematical models of the forces acting on the planets of the solar system recently fit to long data series by a team of astronomers. From the observations, including also three flybys of MESSENGER of Mercury, they
computed the maximum allowed range of values for any unmodeled/mismodeled effect impacting the secular precessions of the planetary perihelia.

If one used them to straightforwardly infer upper bounds on the Sun's angular momentum from a comparison with the theoretical prediction for the Lense--Thirring perihelion precession, it would turn out $S_{\odot}\leq 0.95\times 10^{41}\ {\rm kg\ m^2\ s^{-1}}$. Such a result would be in nice disagreement with the non-dynamical values for $S_{\odot}$ inferred from helioseismology, which, on average, point toward $S_{\odot} = 1.92\times 10^{41}\ {\rm kg\ m^2\ s^{-1}}$.
We discussed some possible explanations. In particular, we looked at several competing dynamical effects that may have conspired to such an outcome. A viable candidate is the residual precession due to the mismodeling in the solar quadrupole mass moment $J_2$, currently known with a $\approx 10\%$ accuracy. Other potential sources of aliasing such as the oblateness of Mercury itself,  the major asteroids, and the ring of minor asteroids are ruled out because their effects are smaller than the current accuracy in determining the orbit of Mercury.

Since the Lense--Thirring perihelion precession of Mercury, as expected from  General Relativity with helioseismology data, is $-2.0\ {\rm mas\ cty^{-1}}$
and the present-day accuracy in constraining the secular perihelion precession of Mercury is  $0.6\ {\rm mas\ cty^{-1}}$ from INPOP10a ephemerides, it seems possible to effectively constrain the solar angular momentum in the near future. To this aim, the Sun's gravitomagnetic force should be explicitly included in the software used by astronomers to reduce data by modifying the coded dynamical force models; it should not be a prohibitive, time-consuming task. As a first step, an analysis of the complete data record from MESSENGER, which was inserted in orbit around Mercury in March 2011 for a year-long science phase extended until March 2013, could be implemented in the next few years. Over a longer timescale, \textit{BepiColombo}, whose launch from the Earth and arrival to Mercury are scheduled for 2015 and 2022, respectively, will further improve our knowledge of the orbit of Mercury allowing for a more refined analysis. It is hoped that  astronomers engaged in the production of modern ephemerides will undertake such an effort, which may be rewarding in no more than ten years.

%% Acknowledgements
%
\begin{acks}
I thank A. Fienga and D. Ragozzine for useful correspondence. I am also grateful to an anonymous referee for all her/his efforts to improve and strengthen the manuscript.
\end{acks}

%-----------------------------------------

\end{article}
\end{document}